\documentclass[aps,prd,preprint,superscriptaddress,tightenlines,nofootinbib]{revtex4}

\usepackage{graphicx}
\usepackage{dcolumn} 
\usepackage{bm}      
\usepackage{epsfig}

\begin{document}

\preprint{CLEO CONF 03-04}
\preprint{EPS 138} 

\title{ Improved Measurement of the Form Factors and First Search for
CP Violation in the Decay $\Lambda_{c}^{+} \rightarrow \Lambda e^+ \nu_e$~}

\thanks{Submitted to the 
International Europhysics Conference on High Energy Physics,
July 2003, Aachen}

\author{Z.~Metreveli}
\author{K.K.~Seth}
\author{A.~Tomaradze}
\author{P.~Zweber}
\affiliation{Northwestern University, Evanston, Illinois 60208}
\author{S.~Ahmed}
\author{M.~S.~Alam}
\author{J.~Ernst}
\author{L.~Jian}
\author{M.~Saleem}
\author{F.~Wappler}
\affiliation{State University of New York at Albany, Albany, New York 12222}
% ...
\collaboration{CLEO Collaboration} 
\noaffiliation

\date{\today}

\begin{abstract}

Using the CLEO detector at the Cornell Electron Storage Ring
we have studied the angular distributions in the decay 
$\Lambda_{c}^{+} \rightarrow \Lambda e^+ \nu_e$.
By performing a four-dimensional maximum likelihood fit, 
we extract the form factor ratio, $R = f_{2}/f_{1}$, and the pole mass, 
which determines the shape of the form factors, $M_{pole}$.
They are found to be $-0.31 \pm 0.05_{stat} \pm 0.04_{sys}$ 
and $(2.13 \pm 0.07_{stat} \pm 0.10_{sys})$~GeV/$c^2$ respectively.
These results correspond to the following value of the decay asymmetry 
parameter $\alpha_{\Lambda_{c}} = -0.85 \pm {0.03}_{stat} \pm 0.02_{sys}$,
for $\langle q^2 \rangle = 0.67$~(GeV/$c^2$)$^2$.
We search for CP violation in the angular distributions of the decay
and find no evidence for CP violation:  $\mathcal{A}_{\Lambda_{c}} = 
\frac{(\alpha_{\Lambda_{c}} + \alpha_{ \overline{\Lambda}_{c} } ) } 
{(\alpha_{\Lambda_{c}} -    \alpha_{ \overline{\Lambda}_{c} } ) } = 
0.01 \pm 0.03_{stat} \pm 0.01_{sys} \pm 0.02_{\mathcal{A}_{\Lambda}}$, 
where the third error  is from  the uncertainty in the world 
average of the CP violating parameter, 
$\mathcal{A}_{\Lambda}$, for $\Lambda \rightarrow p \pi^-$.
All results presented in this paper are preliminary.

\end{abstract}

\maketitle

\collaboration{CLEO Collaboration}


{
\renewcommand{\thefootnote}{\fnsymbol{footnote}}


\setcounter{footnote}{0}
}
\newpage

Charm semileptonic decays allow a measurement of the form factors which 
parameterize the hadronic current because the Cabbibo-Kobayashi-Masawa
(CKM) matrix element $|V_{cs}|$ is known from unitarity~\cite{PDG2000}.  
Within heavy quark effective theory (HQET)~\cite{LambdaInHQET}, 
$\Lambda$-type baryons are more
straightforward to treat than mesons as they consist of a heavy quark 
and a spin zero light diquark.
This simplicity allows for more reliable predictions concerning heavy quark
to light quark transitions~\cite{heavyToLight,KKModel} than is the case for mesons.  
A measurement of the form
factors in $\Lambda_{c}^{+} \rightarrow \Lambda e^+ \nu_e$ 
will help the future determination of the matrix
element $|V_{ub}|$ using $\Lambda_{b}^0$ decays since HQET relates the form factors in 
$\Lambda_{c}^{+}$
decay to those governing $\Lambda_{b}^0$ semileptonic decays.

In the limit of negligible lepton mass, the semileptonic decay of a
charmed
baryon ($1/2^+ \rightarrow 1/2^+$) is usually parameterized in terms of four form
factors:  
two axial form factors $F_{1}^{A}$ and $F_{2}^{A}$ 
and two vector form factors $F_{1}^{V}$ and $F_{2}^{V}$.
These form factors are functions of
$q^2$, the invariant mass squared of the virtual $W^+$.  
In the zero lepton mass approximation, 
the decay may be described in terms of helicity amplitudes
$H_{\lambda_{\Lambda}\lambda_{W}} =
H_{\lambda_{\Lambda}\lambda_{W}}^{V} + H_{\lambda_{\Lambda}\lambda_{W}}^{A}$, 
where $\lambda_{\Lambda}$ and $\lambda_{W}$ are the helicities of the
$\Lambda$ and $W^+$.
The helicity amplitudes are related to the form factors by~\cite{KKModel}

\begin{eqnarray} 
\sqrt{q^2}H^{V}_{\frac{1}{2}0} & = & \sqrt{Q_{-}}\;[(M_{\Lambda_c}+
M_{\Lambda})F^{V}_{1}-q^{2}F^{V}_{2}], \nonumber \\
H^{V}_{\frac{1}{2}1}&=&\sqrt{2Q_{-}}\;[-F^{V}_{1} + (M_{\Lambda_c}+
M_{\Lambda})F^{V}_{2}], \\
\sqrt{q^2}H^{A}_{\frac{1}{2}0} & = & \sqrt{Q_{+}}\;[(M_{\Lambda_c}-
M_{\Lambda})F^{A}_{1}+q^{2}F^{A}_{2}], \nonumber \\
H^{A}_{\frac{1}{2}1}&=&\sqrt{2Q_{+}}\;[-F^{A}_{1} - (M_{\Lambda_c}-
M_{\Lambda})F^{V}_{2}], \nonumber 
\end{eqnarray}

\noindent where $Q_{\pm} = (M_{\Lambda_c} \pm M_{\Lambda})^{2} -q^2$.  
The remaining helicity amplitudes can be obtained using
the parity relations 
$H^{V(A)}_{-\lambda_{\Lambda} -\lambda_{W}} = +(-) H^{V(A)}_{\lambda_{\Lambda}\lambda_{W}}$.  
In terms of the helicity amplitudes, the
decay angular distribution can be written as~\cite{KKModel}~\cite{noteOnPlus}

\begin{widetext}
\begin{eqnarray}
\Gamma && = \frac{d{\Gamma}}{dq^{2}d\cos{\theta_{\Lambda}}d\cos{\theta_W}d\chi}  =   
\mathcal{B}(\Lambda \rightarrow p \pi^-) \frac{1}{2} \frac{G^{2}_{F}}{(2\pi)^4}|V_{cs}|^{2} 
\frac{q^{2}P}{24M^{2}_{\Lambda_c}} \times \nonumber \\  \label{rateTheory} 
&& \{\frac{3}{8}(1 - \cos{\theta_W})^{2}|H_{\frac{1}{2}1}|^{2}(1+
\alpha_{\Lambda}\cos{\theta_{\Lambda}})+ \frac{3}{8}(1+\cos{\theta_W})^2|H_{-\frac{1}{2}-1}|^{2}(1-
\alpha_{\Lambda}\cos{\theta_{\Lambda}}) \\
&& + \frac{3}{4}\sin^{2}{\theta_W}[|H_{\frac{1}{2}0}|^{2}(1+\alpha_{\Lambda}\cos{\theta_{\Lambda}})+ 
|H_{-\frac{1}{2}0}|^{2}(1-\alpha_{\Lambda}\cos{\theta_{\Lambda}})] \nonumber \\ 
&& + \frac{3}{2\sqrt{2}} \alpha_{\Lambda}\cos{\chi}\sin{\theta_W}\sin{\theta_{\Lambda}}[(1 - 
\cos{\theta_{W}})Re(H_{-\frac{1}{2}0}H_{\frac{1}{2}1}^{*})+ \nonumber  
(1+ \cos{\theta_W})Re(H_{\frac{1}{2}0}H_{-\frac{1}{2}-1}^{*})]\}, \nonumber 
\end{eqnarray} 
\end{widetext}

\noindent where $G_F$ is the Fermi coupling constant, $V_{cs}$ is the CKM matrix
element, $P$ is the $\Lambda$ momentum in the $\Lambda^{+}_{c}$
rest frame, $\theta_{\Lambda}$ is the angle
between the momentum vector of the proton in the $\Lambda$  rest frame and
the $\Lambda$  momentum in the $\Lambda^{+}_{c}$ rest frame, 
$\theta_W$ is the angle between the
momentum vector of the positron in the $W^+$ rest frame and the $\Lambda$
momentum in the $\Lambda^{+}_{c}$ rest frame, 
$\chi$ is the angle between the decay planes of the $\Lambda$ and $W^+$,
and $\alpha_{\Lambda}$ is the $\Lambda \rightarrow p \pi^-$
decay asymmetry parameter measured to be $0.642  \pm 0.013$~\cite{PDG2000}.

Within the framework of HQET the heavy flavor and spin symmetries imply 
relations among the form factors which reduce their number to one when
the  decay involves only heavy quarks.  In this Letter we follow Ref.~\cite{KKModel}
where it is argued that the HQET formalism may also work for 
$\Lambda_{c}^{+} \rightarrow \Lambda e^+ \nu_e$.
Treating the $c$ quark as a heavy quark, two independent form factors
$f_1$ and $f_2$ are required to describe the hadronic current.  
The relationships between
these form factors and the standard form factors are 
$F^{V}_{1}(q^2) = -F^{A}_{1}(q^2) = f_{1}(q^2)+\frac{M_{\Lambda}}{M_{\Lambda_c}}f_{2}(q^2)$ 
and 
$F^{V}_{2}(q^2) = -F^{A}_{2}(q^2) = \frac{1}{M_{\Lambda_c}}f_{2}(q^2)$.  
In general $f_2$ 
is expected to be negative and smaller in magnitude than $f_1$.  If the
strange quark is treated as heavy, $f_2$ is zero.

In order to extract the form factor ratio $R = f_{2}/f_{1}$ from a fit to
$\Gamma$ an assumption must be made about the $q^2$ dependence of the
form factors.  
We follow the model of Korner and Kramer (KK)~\cite{KKModel} who use the 
dipole form 
$f(q^2) = \frac{f(q^2_{max})}{(1-q^2/m^2_{pole})^2}(1-\frac{q^2_{max}}{m^2_{pole}})^2$,
where the pole mass is chosen to be $m_{D^{*}_{s}(1^-)} = 2.11$~GeV/$c^2$. 
We will perform a simultaneous fit for the form factor ratio and the pole mass.

The data sample used in this study was collected with the CLEO~II~\cite{cleo2}
and the upgraded CLEO~II.V~\cite{cleo25} detector operating 
at the Cornell Electron Storage Ring (CESR).  
The integrated  luminosity consists of 13.7~fb$^{-1}$ taken at and just below the 
$\Upsilon(4S)$ resonance, corresponding to approximately $18 \times 10^6$
$e^+ e^- \rightarrow c \overline{c}$ events.    Throughout this paper charge
conjugate states are implicitely included, unless otherwise is indicated,
and we use the symbol $e$ to denote an electron or positron.

The analysis technique is an extension of 
previous work~\cite{ffstudy}\cite{lambdacStudy}.
We search for the decay $\Lambda_{c}^{+} \rightarrow \Lambda e^+ \nu_e$ 
in  $e^+ e^- \rightarrow c \overline{c}$ events by detecting a 
$\Lambda e^+$ pair with invariant mass in range 
$m_{\Lambda e^+} < m_{\Lambda_c}$~\cite{ffstudy}.  
All tracks are required to come from the region of the event vertex.  
To reduce the background from $B$ decays, we require $R_2 = H_2 / H_0 > 0.2$, 
where $H_i$ are Fox-Wolfram event shape variables~\cite{FWMoments}.  
Positrons are identified using a 
likelihood function which incorporates information from the calorimeter 
and $dE/dx$ systems.  The minimum allowed momentum for positron 
candidates is 0.7~GeV/$c$, as the positron fake rates are much
higher in the lower momentum range.  Positrons are required to have been 
detected in the region $|\cos{\theta}| < 0.7$, where $\theta$ 
is the angle between the positron momentum and the beam line.

The $\Lambda$ is reconstructed through its decay to $p \pi^-$.  
We require the point of intersection of the two charged tracks, measured in the 
$r - \phi$ plane, to be greater than 5~mm away from the primary vertex.  
In addition, we require the sum of the $p$  and $\pi^-$ momentum 
vectors to extrapolate back to the beam line.  
The $dE/dx$ measurement of the proton is required to be consistent with the expected 
value.  We reject combinations which satisfy interpretation as a $K_{s}^{0}$.  
Finally, we require the momentum of the $p \pi^-$ pair to be greater than 
0.8~GeV/$c$ in order to reduce combinatoric background.  These $\Lambda$
candidates are then combined with Right Sign (RS) tracks consistent
with positrons, and the sum of the $\Lambda$ and $e^+$ momenta
is required to be greater than 
1.4~GeV/$c$, in order to reduce the background from $B$ decays.

The number of events passing these requirements is 4060, of which 
$123 \pm 12$ are consistent with fake $\Lambda$ background,  
$338 \pm 67$ with $\Xi_c \rightarrow \Xi e^+ \nu$ feedthrough
and $398 \pm 58$ are consistent with the $e$ fake background.  
The sidebands of the $p \pi^-$ invariant mass distribution are used to 
estimate the fake $\Lambda$.  The background from $\Xi_c \rightarrow 
\Xi e^+ \nu$ decays is estimated using the result of a previous 
CLEO analysis~\cite{cascadeCStudy}. 
The $e$ fake background is estimated using the Wrong Sign (WS) 
$h^+ \overline{\Lambda}$ (no charge conjugation is implied here) 
data sample, where $h^+ \overline{\Lambda}$ are WS combinations 
passing the analysis selection criteria. 
The $h^+$ tracks in this sample are mostly fakes as there are few 
processes contributing $e^+ \overline{\Lambda}$ pairs after 
the selection criteria are applied. We expect that the estimate
of the normalization and  the momentum spectrum of the $e$ fake 
background from the $h^+ \overline{\Lambda}$ sample is accurate 
for the following reasons. The probability to form 
a $h^+  \Lambda$ or $h^-  \Lambda$ pair is approximately equal 
because the net charge of the event 
is zero. The baryon number and the $(sud)$ quark content of the $\Lambda$ 
imply  (1) that, due to baryon conservation, a $\Lambda$ is more likely to 
be produced with  an antiproton in WS rather than RS combinations and (2) that 
if a $\Lambda$ is produced as a result of a $s \overline{s}$ quark 
pair creation, such events have a higher fraction of kaons in 
RS rather than WS combinations. In addition antiprotons and  kaons have
higher $e$ fake rates. 
By using only one charge conjugate state ($h^+ \overline{\Lambda}$)
and by excluding the momentum region where the $e$ fake
rate from kaons is high by requiring $|\vec{p_e}| > 0.7$~GeV/$c^2$, 
differences between the momentum spectra and particle species
of hadronic tracks  between $h^+  \overline{\Lambda}$ and $h^+  \Lambda$ 
are minimised. Differences that remain are second order and are 
accounted for as a systematic uncertainty on the final result.

Calculating kinematic variables requires knowledge of the $\Lambda_{c}^+$  
momentum which is unknown due to the undetected neutrino.  The direction 
of the $\Lambda_{c}^+$ is 
approximated based on the information provided by the thrust axis of the event
and the kinematic constraints of the decay.
The  magnitude of the $\Lambda_{c}^+$ momentum is obtained as a weighted average of the roots
of the quadratic equation $\vec{p}^{~2}_{\Lambda_c} = (\vec{p}_{\Lambda} + \vec{p}_{e} + 
\vec{p}_{\nu})^2$.  The weights  are assigned
based on the fragmentation function of $\Lambda_{c}^{+}$.
After the $\Lambda_{c}^{+}$ momentum is estimated, the four kinematic variables are
obtained  by working in the $\Lambda_{c}^{+}$ center-of-mass frame.

Using $t = q^2/q^{2}_{max}$, $\cos{\theta_{\Lambda}}$, $\cos{\theta_{W}}$, and $\chi$, 
we perform a four-dimensional maximum 
likelihood fit in a manner similar to Ref.~\cite{fitMethod}.  
The technique enables a multidimensional likelihood fit to be performed 
to variables modified by experimental acceptance and resolution and 
is necessary for this analysis 
due to the substantial smearing of the kinematic variables.  The essence 
of the method is to determine the probability density function by using 
the population of appropriately weighted Monte Carlo events in the 
four-dimensional kinematic space.  This is accomplished by generating one 
high statistics sample of Monte Carlo events with known values of the form
factor ratio $R$, the pole mass $M_{pole}$ and corresponding known values of the four 
kinematic variables $t$, $\cos{\theta_{\Lambda}}$, $\cos{\theta_{W}}$, 
and $\chi$ for each event.  The generated events are then processed 
through the full detector simulation, off-line analysis programs, and 
selection criteria.  Using the generated kinematic variables, the accepted 
Monte Carlo events are weighted by the ratio of the decay distribution for the 
trial values of $R$ and $M_{pole}$ to that of the generated distribution.  
The accepted Monte Carlo events are now, therefore, distributed according to the probability 
density corresponding to the trial values of $R$ and $M_{pole}$.  
By such weighting, a likelihood may be evaluated for each data event for 
different values of the form factor ratio and the pole mass, 
and a fit performed.  

The fit is unbinned in $\cos{\theta_{\Lambda}}, \cos{\theta_W}$ and $\chi$ and takes 
into account the correlations among these variables.
The probability for each event is determined by sampling this distribution using a 
search volume around each data point.  The volume size is chosen so that the
systematic effect from finite search volumes is small and the required number of Monte Carlo 
events is not prohibitively high. 
We bin the fit in $t$ for the following reason. 
In the course of this analysis we observed an 
inconsistency between datasets for
one subsample of data for $0.8<t<1.0$ 
(a total of 237 evnets). The projections in the other
three variables for this subsample are 
consistent with the remainder of the data.
To avoid bias we wish to include these events 
in the fit for three variables, but not for $t$.
This is not possible if the fit is unbinned in 4 dimensions. 
Instead we include the events in the 3 dimensional fit but, 
by virtue of perfoming a binned fit in $t$, we 
exclude their contribution to the calculation of 
the likelihood in $t$. A systematic uncertainty is assigned by varying 
the excluded region.

The $\Lambda_{c}^+ \rightarrow \Lambda e^+ \nu$ sample has signal to 
background in the approximate ratio 4:1.  Background is incorporated into the 
fitting technique by constructing the log-likelihood function 

\begin{widetext}
\begin{eqnarray}
{\mathcal{L}} = 
 \{ \sum_{i=1}^{N_{events}} 2\ln{(P_{S}^{i}{\Gamma}_{S}^{i} +
\sum_{j} P_{j~B}^{i}{\Gamma}_{j~B}^{i})} \}
- \{\sum_{i=1}^{N_{bins}} {(n_{observed}^i -  
n_{estimated}^i)^2}/{\sigma^{2}_{n_{observed}^i}}\}.
\end{eqnarray}
\end{widetext}

\noindent The first term  is the sum over the number of events, where $N_{events}$ is the 
number of events in the signal region, $P^{i}_{S}$ and $P^{i}_{j~B}$  
are the probabilities for the $i$th event to be signal and belong to the
$j$th background component, respectively, 
${\Gamma}^{i}_{S}$ is  the signal shape modeled from signal Monte Carlo, 
and ${\Gamma}^{i}_{j~B}$ is the background shape modeled by a
sample of events for the $j$th background component.
The second term is the sum  over bins of $t$, where
$n_{observed}^i$ is the number of events in the $i$th bin, $n_{estimated}^i$ is
our prediction of that number, which is dependent on the trial values 
for $R$ and $M_{pole}$ and is estimated
similarly to ${\Gamma}^{}_{S}$ and  
${\Gamma}^{i}_{j~B}$.

The $e$ fake background is modeled by the fake positron data sample.  
Feedthrough background from $\Xi_c \rightarrow \Xi e^+ \nu$ is modeled 
by the Monte Carlo sample which is generated according to the 
HQET-consistent KK model.  Fake $\Lambda$ background is modeled using
the data  events in the sidebands of the $p \pi^-$ invariant mass 
distribution.

\begin{figure}
\begin{center}
      \epsfig{figure=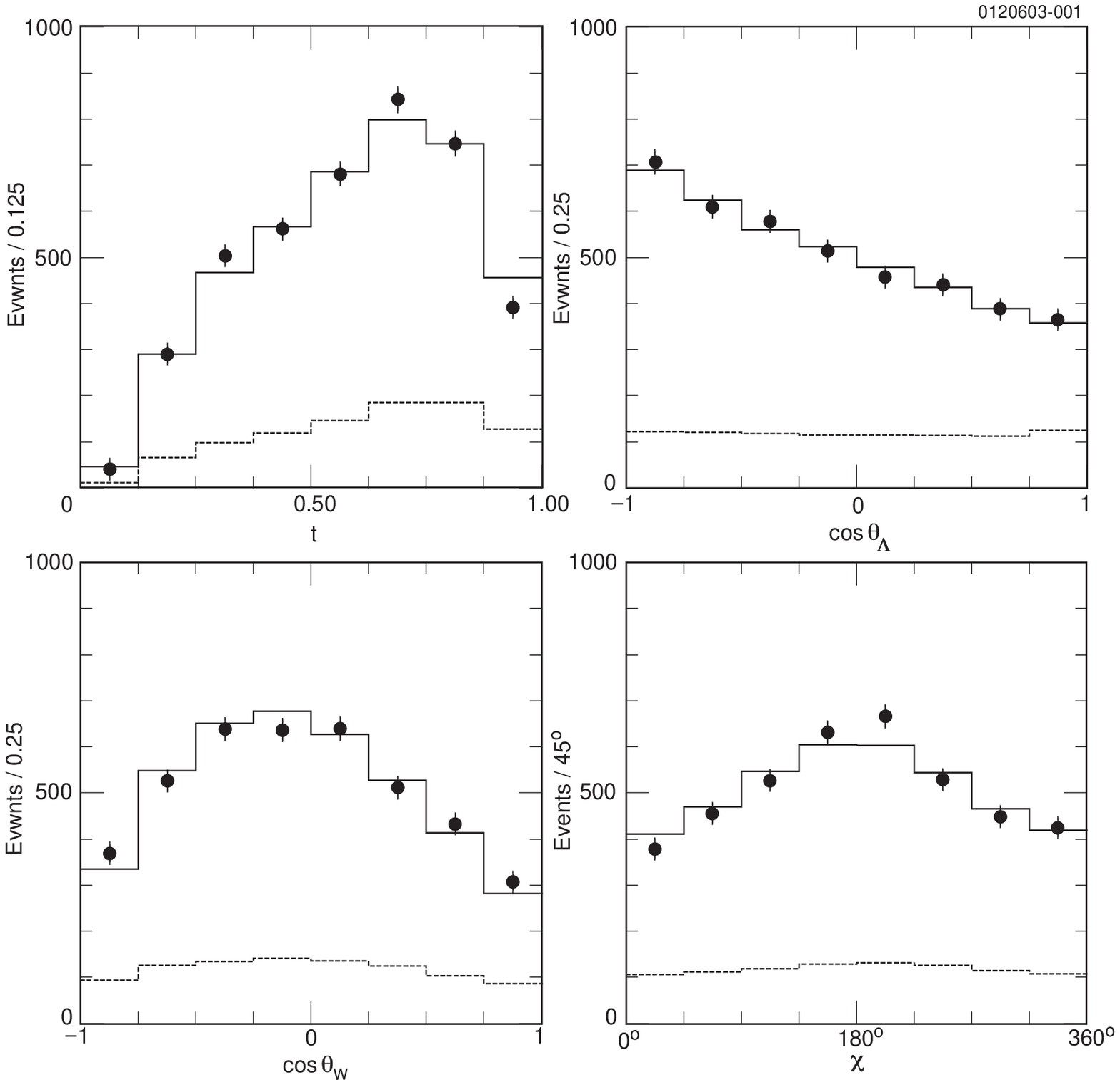,height=3.0in}
\end{center}
   \caption{ Projections of the data (points with error bars) and the fit
(solid histogram) onto $t$, $\cos\theta_{\Lambda}$, $\cos\theta_W$ and $\chi$.
The dashed lines show the sum of the background distributions. }
   \label{projections1}
\end{figure}

\begin{figure}
\begin{center}
      \epsfig{figure=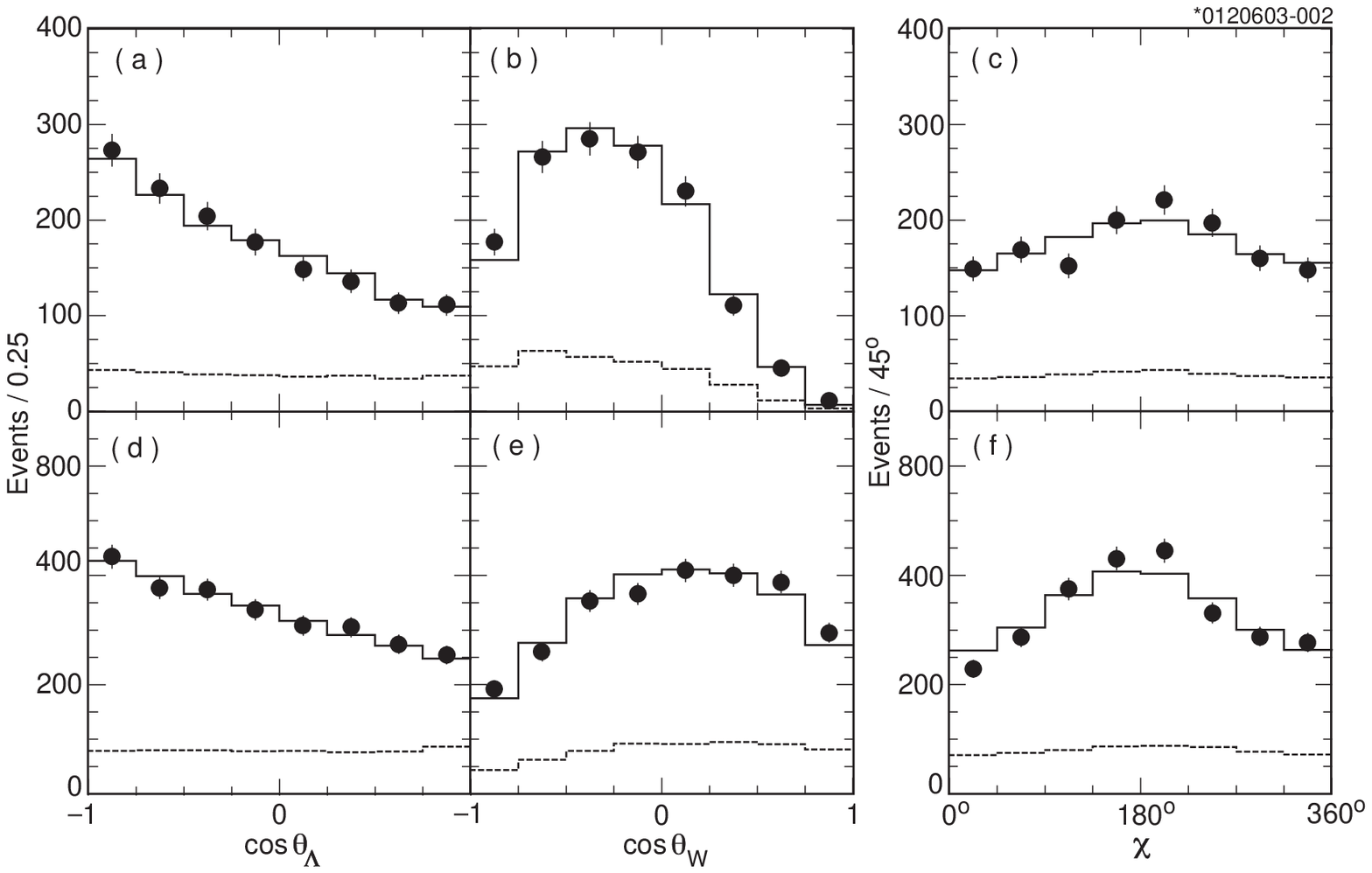,height=3.0in}
\end{center}
   \caption{ Projections of the data (points with error bars) and the fit
(solid histogram) onto $\cos\theta_{\Lambda}$, 
$\cos\theta_W$ and $\chi$  for two $t$ regions. 
The plots labeled (a), (b) and (c) are for $t < 0.5$;
(d), (e) and (f) are for  $t > 0.5$.
The dashed lines show the sum of the background distributions.
}
\label{q2BinsProjections}
\end{figure}

Using the above method we perform a simultaneous fit for 
the form factor ratio and the pole mass and  find 
$R = -0.31 \pm 0.05$ and $M_{pole} = (2.13 \pm 0.07)$~GeV/$c^2$,
where the uncertainties are statistical. This is our main result. 
Figures~\ref{projections1} and~\ref{q2BinsProjections} show the 
$t$, $\cos\theta_{\Lambda}$, $\cos\theta_W$ and $\chi$ projections 
for the data and for the fit.

We have considered the following sources of systematic uncertainty and give
our estimate of their magnitude in parentheses for $R$ and $M_{pole}$.
The uncertainty associated with the size of the search volume is measured
from a statistical experiment in which a set of mock data samples,
including signal and all background components,
was fit in the same way as the data~(0.006, 0.048). 
The uncertainty due to the limited size of the Monte Carlo sample 
is estimated by dividing the Monte Carlo sample
into four independent equal samples and repeating the fit (0.007, 0.012).
The uncertainty due to background normalizations is determined by varying 
estimated number of the background events in the signal region by one 
standard deviation for each type of background separately~(0.023, 0.024).
The uncertainty associated with the modeling of the background
shapes,  including uncertainties originating from the unknown form factor ratio
and pole mass for the decay $\Xi_c \rightarrow \Xi e^+ \nu$, 
are estimated by varying these shapes and by using 
different background samples (0.024, 0.049).
The uncertainty from random $e^+ \Lambda$ pairs from in 
$q \overline{q}$ and $\Upsilon(4S) \rightarrow B \overline{B}$ events
is estimated by repeating the fit with and without correcting for
this background contributions (0.013, 0.038).
The modes  $\Lambda^{+}_c \rightarrow
\Lambda X e^+ \nu$,  where $X$ represents unobserved decay products,
have never been observed. The current upper limit is
$\mathcal{B}(\Lambda^{+}_c \rightarrow
\Lambda X e^+ \nu)/\mathcal{B}(\Lambda^{+}_c \rightarrow
\Lambda e^+ \nu) < 0.15$, where $X \ne 0$, at 90\% confidence level~\cite{lambdacStudy}.
The uncertainty due to the possible presence of these modes
is estimated from a series of fits, each with an additional background component
representing  possible $\Lambda^{+}_c \rightarrow \Lambda X e^+ \nu$ modes. 
The normalizations of the additional components are allowed to float in the fits.
By performing fits with a variety of $\Lambda^{+}_c \rightarrow
\Lambda X e^+ \nu$ modes, and combining the results, 
we obtain an estimate of the
uncertainty in the final result due to the possible
presence of $\Lambda^{+}_c \rightarrow
\Lambda X e^+ \nu$~(0.020, 0.060).
The uncertainty associated with the $\Lambda_{c}^+$
fragmentation function is estimated by varying this function (0.003, 0.002). 
The uncertainty associated with Monte Carlo modeling of slow pions from $\Lambda$ 
decay is obtained by varying this efficiency according to our understanding of
the  CLEO detector (0.004, 0.003).  
Finally, we account for the effect of excluding $0.8<t<1.0$  by
varying the size of the excluded region~(0.000; 0.014).
Adding all sources of systematic uncertainty in quadrature, our final result
is  $R = -0.31 \pm 0.05_{stat} \pm 0.04_{sys}$ and 
$M_{pole} = (2.13 \pm 0.07_{stat} \pm 0.10_{sys})$~GeV/$c^2$.
We also find $R = -0.32 \pm 0.04_{stat} \pm 0.03_{sys}$ for
a fit for the form factor ratio with 
$M_{pole} = 2.11$~GeV/$c^2$ fixed.

Using the value of $R$ and $M_{pole}$ obtained in the 
simultaneous fit and the HQET-consistent KK model, 
the mean value of the decay asymmetry parameter of 
$\Lambda_{c}^{+} \rightarrow \Lambda e^+ \nu_e$ 
averaged over the charge conjugate states is calculated to be 
$\alpha_{\Lambda_{c}} =-0.85 \pm {0.03}_{stat} \pm 0.02_{sys}$,
for $\langle q^2 \rangle = 0.67$~(GeV/$c^2$)$^2$.

In the Standard Model CP violation is expected to be very small
for semileptonic decays. 
In the KK model, the shape of the decay rate 
distribution in the 4D space of $t$, $\cos{\theta_{\Lambda}}$, $\cos{\theta_W}$ and 
$\chi$ is governed by $R$ and $M_{pole}$ only. 
We determine $\alpha_{\Lambda_{c}} \alpha_{\Lambda}$ 
and $\alpha_{\overline{\Lambda}_{c}} \alpha_{\overline{\Lambda}}$ by 
repeating the simultaneous fit
to $R$ and $M_{pole}$ for each charge conjugate state separately.
We find $\alpha_{\Lambda_{c}} \alpha_{\Lambda} = -0.561 \pm 0.026_{stat}$ and 
$\alpha_{\overline{\Lambda}_{c}} \alpha_{\overline{\Lambda}} =  
-0.535 \pm 0.024_{stat} $.
Following~\cite{acp} and by extension 
we define the CP violating asymmetry of the $\Lambda^+_c$ as
$\mathcal{A}_{\Lambda_{c}} =  \frac {(\alpha_{\Lambda_{c}} + 
\alpha_{\overline{\Lambda}_{c}})}  {(\alpha_{\Lambda_{c}} - 
\alpha_{\overline{\Lambda}_{c}})}$.
From the measurement of the CP asymmetry in
the product of $\alpha_{\Lambda_{c}}\alpha_{\Lambda}$, 
and using the relation
\begin{eqnarray}
\frac{\alpha_{\Lambda_{c}}\alpha_{\Lambda}  - 
\alpha_{\overline{\Lambda}_{c}} \alpha_{\overline{\Lambda}}}
{\alpha_{\Lambda_{c}}\alpha_{\Lambda}  + 
\alpha_{\overline{\Lambda}_{c}}\alpha_{\overline{\Lambda}}} 
= \mathcal{A}_{\Lambda_{c}} + \mathcal{A}_{\Lambda} + 
\mathcal{O}(\mathcal{A}_{\Lambda_{c}}^{2},
\mathcal{A}_{\Lambda}^{2}),
\end{eqnarray}
\noindent we obtain $\mathcal{A}_{\Lambda_{c}} = 
0.01 \pm 0.03_{stat} \pm 0.01_{sys} \pm 0.02_{\mathcal{A}_{\Lambda}}$, 
where in the systematic uncertainty
we have taken into account the correlations among the systematic 
uncertainties for the charge conjugate states and
the third error is from the uncertainty in the world average
of the CP violating parameter, $\mathcal{A}_{\Lambda}$, 
for $\Lambda \rightarrow p \pi^-$~\cite{PDG2000}.

In conclusion, using a four-dimensional maximum likelihood fit the 
angular distributions of $\Lambda_{c}^{+} \rightarrow \Lambda e^+ \nu_e$ 
have been studied and the form factor ratio $R = f_2 / f_1$ and $M_{pole}$
are found to be $-0.31 \pm 0.05_{stat} \pm 0.04_{sys}$ and 
$(2.13 \pm 0.07_{stat} \pm 0.10_{sys})$~GeV/$c^2$ respectively. 
These results correspond to the following value
of the decay asymmetry parameter
$\alpha_{\Lambda_{c}} =-0.85 \pm {0.03}_{stat} \pm 0.02_{sys}$,
for $\langle q^2  \rangle = 0.67$~(GeV/$c^2$)$^2$.
We have searched for CP violation in the angular distributions of the 
decay and find no evidence for CP violation:  $\mathcal{A}_{\Lambda_{c}} = 
0.01 \pm 0.03_{stat} \pm 0.01_{sys} \pm 0.02_{A_{\Lambda}}$, 
where the third error is  from  the uncertainty in the 
world average of the CP violating parameter, 
$\mathcal{A}_{\Lambda}$, for $\Lambda \rightarrow p \pi^-$.
All results presented in this paper are preliminary.

We gratefully acknowledge the effort of the CESR staff 
in providing us with excellent luminosity and running conditions.
This work was supported by  the National Science Foundation,
the U.S. Department of Energy, the Research Corporation,
and the  Texas Advanced Research Program.


\begin{thebibliography}{99}

\bibitem{PDG2000}
Particle Data Group, D.E. Groom {\it et al.,} Eur. Phys. J. C. {\bf 15}, 1 (2000).

\bibitem{LambdaInHQET} 
N. Isgur and M.B. Wise, Phys. Lett. B {\bf 232}, 113 (1989); {\bf 237}, 527 (1990); 
E. Eichten and B. Hill, Phys. Lett. B {\bf 234}, 511 (1990); 
H. Georgi, Phys. Lett. B {\bf 240}, 447 (1990).

\bibitem{heavyToLight}  
T. Mannel, W. Roberts, and Z. Ryzak, Nucl. Phys. {\bf B355}, 38 (1991); 
Phys. Lett. B {\bf 255}, 593 91991); 
F. Hussain, J.G. Korner, M. Kramer, and G. Thompson, Z. Physics C {\bf 51}, 321 (1991); 
A.F. Falk and M. Neubert, Phys. Rev D {\bf 47}, 2982 (1993); 
H. Georgi, B. Grinstein and M.B. Wise, Phys. Lett. B {\bf 252}, 456 (1990); 
N. Isgur and M.B. Wise, Nucl. Phys. {\bf B348}, 276 (1991); 
H. Georgi, Nucl. Phys. {\bf B348}, 293 (1991).


\bibitem{KKModel} J.G. Korner and M. Kramer, Phys. Lett. B {\bf 275}, 
495 (1992).

\bibitem{noteOnPlus}


The interference term (the term containing $\cos{\chi}$) in expression~(\ref{rateTheory}) 
contributes with a positive sign to the decay angular distribution. In the data, the sign of the
term is opposite (Figure~\ref{projections1}). 
In our derivation we reproduced the expressions for the helicity amplitudes in terms of
the traditional form factors and the expressions relating  the traditional form 
factors to the two HQET consistent form factors, which are given in~\cite{KKModel}. 
We used the helicity formalism to derive the four fold decay rate and to reproduce all 
the terms and their signs; the only exception was the sign of the interference term. 
In the rest of the analysis it is assumed that the correct sign of the interference 
term is minus.

\bibitem {cleo2} Y. Kubota {\it et al.,} Nucl. Instrum. Methods 
Phys. Res. Sect. A {\bf 320}, 255 (1992).



\bibitem{cleo25} T. Hill, Nucl. Instrum. Methods 
Phys. Res. Sect. A {\bf 418}, 32 (1998).


\bibitem{ffstudy}
CLEO Collaboration, 
G.L. Crawford  {\it et al.}, Phys. Rev. Lett. {\bf 75}, 624 (1995).


\bibitem{lambdacStudy}  CLEO Collaboration, T. Bergfeld {\it et al.}, Phys. Lett. B {\bf 323}, 
219  (1994).


\bibitem{FWMoments}
    G.C.Fox and S.Wolfram $et\;el.$, Phys. Rev. Lett. {\bf 41} 1581 (1978).



\bibitem{cascadeCStudy}  CLEO Collaboration,  
J. Alexander $et~al.$, Phys. Rev. Lett. {\bf 74}, 3113 (1995).

\bibitem{fitMethod}  D. M. Schmidt, R.J. Morrison and M.S. Witherell, Nucl. Instrum. 
Methods Phys. Res., Sect. A {\bf 328}, 547 (1993).


\bibitem{acp}
F. Donoghue, and S. Pakvasa, Phys. Rev. Lett. {\bf 55},  
162 (1985).


\end{thebibliography}
\end{document}